\documentclass[english,showpacs]{revtex4-1}
\usepackage[T1]{fontenc}
\usepackage[latin9]{inputenc}
\usepackage{textcomp}
\usepackage{amsmath}
\usepackage{amssymb}
\usepackage{esint}
\usepackage{epsfig,latexsym,bm,longtable}
\usepackage{multirow}

\usepackage{latexsym}

\makeatother

\usepackage{babel}
\begin{document}

%
%
%
%
%
%
%
%
%

\title{K-essence scalar field as dynamical dark energy}

\author{L. A. Garc\'ia}
\email{lagarciape@unal.edu.co}
  
\author{J. M. Tejeiro}
\email{jmtejeiros@unal.edu.co}

\author{L. Casta\~neda}
\email{lcastanedac@unal.edu.co}
\affiliation{Observatorio Astron\'omico Nacional, Universidad Nacional de Colombia.}

%

\begin{abstract}
We study an early dark energy (EDE) model as a K-essence scalar field in the framework of FLRW universe 
using an effective parametrization of the state equation as a function of the redshift $z$ with the tracker 
condition during radiation domination, but also demanding an accelerated expansion of the universe at late times emulating cosmological constant. We found all the dynamical variables of the EDE system. We use the luminosity distances of the SNIA to get the best estimations for the free parameters of the model and also, we constrain the model using primordial abundances of light nuclei in BBN theory. We summarize the necessary conditions to achieve BBN predictions and the accelerated expansion of the universe at late times. \\

\end{abstract}

\maketitle

\section{\label{sec:level1}Introduction}

Recent observations of the luminosity distances of the SNIa \cite{perlmutter} reveal that the expansion of the universe is accelerated and there is an unknown matter--energy contribution as about 70\% of the critical density, which is smooth and has negative pressure. In order to explain this phenomenon, it has been proposed many plausible solutions: the cosmological constant $\Lambda$ which is related to the vacuum energy of the quantum fields \cite{carroll}, Quintessence fields (with the state equation $\omega=\frac{p_{Q}}{\rho_{Q}}$=constant), Kessence, Taquionic fields, frustrated topological defects, extra--dimensions, massive (or massless) fermionic fields, galileons, effective parametrizations of the state equation, primordial magnetic fields, holographic models, etc. All these proposals have been used to model this contribution predicted by the Friedmann equations in the framework of the General Relativity. Also, there are other possible options as Modified Gravity, where the accelerated expansion effect is geometric and not as a matter--energy form.\\
Nowadays, the current paradigm is the $\Lambda$CDM model, which is so far the best fitting to the present observations, even though the conceptual problems that persist with the nature of $\Lambda$.\\

On the other hand, primordial abundances of the light nuclei that were formed during Big Bang Nucleosynthesis (\textbf{BBN}) are well observed and quantified with astrophysical methods, specially the mass fraction of the $^4He$. Actually, there is an extended theory of BBN, introduced by Alpher, Bethe and Gamov \cite{alpher, gamov}, and it has been developed many numerical codes that resolve the Boltzmann equation for each isotope including the cosmological background (based on Kawano code). 
There is a good agreement between the predicted abundances and the observations, however, the main problem remains with $^7Li$. Despite the efforts that have been made, including corrections on the cross sections of this element, the discrepancies are not negligible.\\

With the aim to enhance the primordial abundances calculated from BBN and also, give a plausible explanation of the accelerated expansion of the universe, we have proposed a model of dark energy which has a non--null contribution at early times to increase the Hubble radius during radiation domination era and influence the Boltzmann equations that determine the evolution of the light abundances. All the conditions that allow us to describe the early dark energy are achieved with a K-essence scalar field, which is characterized by its state equation that overcomes the attractors defined by a dynamical system.\\

In the first section, we expose the main conditions imposed to the scalar field and we resolve the dynamical system that appears from the cosmological assumptions. At the second part, we propose the effective parametrization of the state equation, determinate the best estimations of the free parameters of the model using the luminosity distances of the SNIa from the Union dataset and derive analytical expressions for the dynamical variables of the K-essence system. In the IV section, we test our model with some standard proves and summarize the main differences between this model and $\Lambda$CDM.\\

Taking into account that there is a non-negligible energy density of the K-essence field during radiation era, we compute the BBN abundances, including the scalar field degrees of freedom in Hubble parameter, following the derivation and discussion made by Bernstein \cite{bernstein2} and using the available codes for Nucleosynthesis. Finally, we find the values of $\Omega_B$ and $\eta_B$ predicted by our model as a result of the BBN calculations.

\section{K-essence scalar field and the dynamical system}

The K-essence scalar fields appeared at the late 90s with the K-inflation model proposed by Armendariz-Picon \cite{armendariz1, armendariz2}. However, the idea was extended to describe a dynamical dark energy contribution, taking into account that this field can track during radiation domination epoch and also, it could avoid the fine--tuning of the initial values of the field and its velocity. These features are well known for the system, so the challenge is resolve the evolution equations of the field in the FLRW spatially flat universe and without cosmological constant.\\

The lagrangian of the K-essence scalar field is given by:

\begin{equation}\label{1}
p(X,\phi)=K(\phi)L(X),
\end{equation}

\noindent where $X=-\frac{1}{2}\nabla^{\alpha}\phi\nabla_{\alpha}\phi$ is the kinetic energy of the field and $v$ its velocity $v=\frac{d\phi}{dt}=\sqrt{-2X}>0$. Rewriting \eqref{1} in terms of $v$, the lagrangian has the following form :

\begin{equation}\label{2}
p(v,\phi)=K(\phi)Q(v).
\end{equation}

\noindent The most general action that describes the K-essence field in a cosmological plasm is given by:

\begin{equation}\label{accion}
S=\int d^4x \sqrt{-g}\left(\frac{R}{2\kappa ^2}+p(\phi,X)\right)+S_B,
\end{equation}

\noindent with $\kappa ^2= 8 \pi G$ and $S_B$ is the action of the background matter. The signature that is used is $\{-1,+1,+1,+1\}$ The equations of motion of the field come from the variation of the lagragian with respect to the field $\phi$:

\begin{equation}\label{eqmov}
\left(\frac{1}{v}\frac{\partial p}{\partial v}+v\frac{\partial}{\partial v}\left(\frac{1}{v}\frac{\partial p}{\partial v}\right)\right) \ddot{\phi}+\frac{\partial p}{\partial v}(3H) +v\frac{\partial ^2 p}{\partial \phi \partial v}-\frac{\partial p}{\partial \phi}=0.
\end{equation}

\noindent From \eqref{accion}, it is obtained the energy--momentum tensor:

\begin{equation}
T_{\mu \nu}=\partial_{\mu}\phi\partial_{\nu}\phi-p(\phi, v)g_{\mu \nu}.
\end{equation}

\noindent Since the scalar field can be described as a perfect fluid, the energy density and the pressure are defined by:

\begin{equation}\label{rho}
\rho_{\phi}=K(\phi)\left(v\frac{\partial Q}{\partial v}-Q\right),
\end{equation}

\begin{equation}\label{p}
p_{\phi}=K(\phi)Q(v).
\end{equation}

\noindent In addition, the adiabatic velocity of sound for the K-essence field is given by:

\begin{equation}\label{cs}
C_s^2=\dfrac{Q^{\prime}}{v Q^{\prime\prime}}.
\end{equation}

\noindent where $\prime$ denotes derivative with respect to the velocity of the field $v$. $C_s^2$ gives relevant information of the stability of the perturbations associated with the K-essence field.\\ 
In order to resolve the \eqref{eqmov} and find an explicit form of \eqref{rho} and \eqref{p}, there have been suggested many alternatives: fixing a specific function of $\phi(t)$ or $v(t)$ \cite{chiba}, making  redefinition of the field to face a modified lagrangian $\bar{Q}(v)$ \cite{copeland}, considering pure kinetic K-essence model \cite{stiff, linder, nojiri, yang, scherrer2} or imposing Slow roll conditions on $K(\phi)$. \\\\

However, we want to resolve the complete dynamical system defined by the Friedmann and the continuity equations for non-interactive fluids in the cosmological background: a matter (or radiation component) and the K-essence scalar field. 

\begin{equation}\label{ec2}
H=\frac{\dot{a}}{a}=\kappa \sqrt{\rho_{\phi}+\rho_m},\:\:\:\:\:\:  
\end{equation}
\begin{equation}\label{ec3}
\frac{\ddot{a}}{a}=-\frac{\kappa^2}{6}((1+3\omega_{\phi})\rho_{\phi}+(1+3\omega_{m})\rho_m), \:\:\:\:\:\: 
\end{equation}
\begin{equation}\label{ec4}
\dot{\rho_m}=-3H(1+\omega_{m})\rho_m, \:\:\:\:\:\:\:\:\:\:\:\:  \dot{\rho_{\phi}}=-3H(1+\omega_{\phi})\rho_{\phi},\:\:\:\:\:\:  
\end{equation}

\noindent Reexpressing the equiations \eqref{ec2}, \eqref{ec3}, \eqref{ec4} in terms of the velocity and the dimensionless variable $F=\frac{\rho_{m}}{\rho_{\phi}+\rho_{m}}$ (matter energy density fraction):
\begin{equation}\label{bep}
\frac{dv}{d\phi}=-C_s^2\left[ \frac{(lnK)_{,\phi}v}{1+\omega_{\phi}}+3\kappa \left(\frac{K \tilde{\rho_{\phi}}}{1-F}\right)^{1/2} \right],
\end{equation}

\begin{equation}\label{bip}
\frac{dF}{d\phi}=-\frac{3\kappa}{v}F\sqrt{1-F}\sqrt{K \tilde{\rho_{\phi}}}\left(\omega_{m}-\omega_{\phi}\right).
\end{equation}

\noindent In addition, it must be fulfilled the following conditions:

\begin{equation}\label{bo}
\frac{d\tilde{\rho_{\phi}}}{dv}=\frac{(1+\omega_{\phi})}{v C_s^2}\tilde{\rho_{\phi}},
\end{equation}

\begin{equation}\label{bu}
\frac{d\omega_{\phi}}{dv}=\frac{1+\omega_{\phi}}{v}\left( 1-\frac{\omega_{\phi}}{C_s^2}\right).
\end{equation}

\noindent It is assumed an asymptotic behaviour for the function $K(\phi)$ \cite{kang}:

\begin{equation}\label{k}
K(\phi)=\frac{1+K_0(\phi)}{\phi^2}, \:\:\:\:\:\:\:\:\: \lim_{\phi \to \infty} K_0(\phi)=0.
\end{equation}

\noindent To reach an attractor during radiation domination epoch, the dynamical system $\{v(\phi),R(\phi)\}$ must fulfill the \textit{ansatz}:

\begin{equation}\label{sol}
v(\phi)=v_0-A(\phi), \:\:\:\:\:\:\:\:\: F(\phi)=F_0-B(\phi).
\end{equation}

\noindent where $A(\phi),B(\phi)\rightarrow 0$ monotonically for $\phi \to \infty$ (or equivalently, $v=v_{rad}$ and $F\sim 0$). The ansatz implies the following physical conditions on the system:

\begin{equation}\label{at1}
\omega_m=\omega_{\phi}(v_0), \:\:\:\:\:\:\:\:  \tilde{\rho_{\phi}}(\phi)\neq 0, \:\:\:\:\:\:\:\:  C_s^2(v_0)>\omega_m
\end{equation}
\begin{equation}\label{at2}
\omega_{\phi}^{\prime}(v_0)=\left(1-\frac{\omega_m}{C_s^2(v_0)}\right)\frac{1+\omega_m}{v_0}\neq 0.
\end{equation}

\noindent There is also a De-Sitter attractor given by the condition $R\sim 0$, which guarantees the existence of a accelerated expansion at late times:

\begin{equation}\label{sol2}
v(\phi)=v_s-A(\phi), \:\:\:\:\:\:\:\:\: R(\phi)=B(\phi).
\end{equation}

\noindent The last ansatz entails a condition on the state equation on the vicinity of the De-Sitter attractor:

\begin{equation}\label{timon}
\dfrac{{1-\omega_{\phi}(v_s)}}{{1+\omega_{\phi}(v_s)}}>0,\:\:\:\:\:\:\:\:  \vert \omega_{\phi}(v_s) \vert <1.
\end{equation}

\section{Effective parametrization of $\omega_{\phi}$}

In order to get a general solution for the dynamical system  defined by \eqref{bo}, \eqref{bu}, \eqref{at1}, \eqref{at2} and \eqref{timon}, it has been proposed an effective parametrization of the state equation from $z<10^{15}$ given by:

\begin{equation}\label{fun}
\omega_{\phi}(z)=\frac{4/3}{\left(\frac{1+z_d}{1+z}\right)^m+1}-1,
\end{equation}

\noindent where $m$ is factor that modules the transitions between the attractors, $z_d$ is a redshift in matter domination epoch defined by $z_{d}=\dfrac{z_{eq}+z_{*}}{2}$ and $z_*$, the redshift where the De-Sitter domination -accelerated expansion- begins.\\

The parametrization \eqref{fun} respects all the conditions previously mentioned, hence it is possible to resolve the functions related to the K-essence lagrangian. \\
The energy density of the field $\rho_{\phi}$  \eqref{rho} results:

\begin{equation}\label{intrho}
\int_{\rho}^{\rho_0} \frac{d\rho^{\prime}}{\rho^{\prime}}=-3\int_{a}^{1} \frac{(1+\omega_{\phi}(a^{\prime}))}{a^{\prime}}da^{\prime},
\end{equation}

\noindent integrating \eqref{intrho}, it is obtained:

\begin{equation}\label{rhoend}
\rho=\rho_0 \cdot (1+z)^{4} \left[\frac{\left(\left(\frac{1+z_d}{1+z}\right)^m+1\right)} {\left(\left(1+z_d\right)^m+1\right)} \right]^{4/m}=\rho_0 \cdot f(z).
\end{equation}

\noindent with 
\begin{align}\label{fa}
f(a)&=exp\left[ -3\int_{a}^{1} \frac{(1+\omega_{\phi}(a^{\prime}))}{a^{\prime}}da^{\prime} \right] \\
&= a^{-4} \left[\frac{\left(\left(\frac{a}{a_{d}}\right)^m+1\right)} {\left(\left(\frac{1}{a_{d}}\right)^m+1\right)} \right]^{4/m},
\end{align}

\noindent Meanwhile, the fraction of the dark energy density $\Omega_{\phi}=1-F=\frac{\rho_{\phi}}{\rho_{cr}}$:

\begin{equation}\label{frack}
\Omega_{\phi}=\frac{\Omega_{\phi 0} \cdot f(a)}{\Omega_{\phi 0}\cdot  f(a) + \Omega_{m 0}\cdot  a^{-3}},
\end{equation}

The formal solution of \eqref{fun} will be obtained with the best estimation
of the free parameters of the model $\{\Omega_{\phi_0}, m, z_{*}\}$. For this
reason, we use the luminosity distances of the SNIa from the survey
\textit{Supernova Cosmology Project} 
 with $z>0.8$ to minimize the function $\chi ^2$:

\begin{equation}\label{chisquare}
\chi ^2= \sum _{i=1}^{N \sim 59} \frac{[\mu_{i}-\mu(z_i)]^2}{\sigma^2}
\end{equation}

\noindent with the distance modulus given by the expression $\mu=m-M= 5 (log_{10}d_L(z)-1)$ and the luminosity distance of our model:

\begin{equation}\label{distance}
d_L (z)= \frac{c (1+z)}{H_0}\int_{0}^{z}\frac{dz^{\prime}}{B(z)}.
\end{equation}

\begin{equation*}
B(z)=(\Omega_{\phi_0}f(z^{\prime};m, z_{*})+(1-\Omega_{\phi_0})(1+z^{\prime})^{3})^{1/2}
\end{equation*}

\noindent \eqref{chisquare} must be resolved together with the constraints:

\begin{equation}\label{omega}
\sum _{i=1}^{N \sim 59} \left(\frac{\mu_{i}-\mu(z_i)}{\sigma^2}\right)\left(\frac{\partial \mu(z_i;\Omega_{\phi_0},m, z_{*})}{\partial x}\right)=0, 
\end{equation}

\noindent where $x=\Omega_{\phi_0},m, z_{*}$. In addition, we have imposed other 2 conditions for the parameters: the deceleration parameter has to be zero at $z_*$, therefore $q(z_*)=0$, but also $f(z\sim z_{BBN}\approx 10^9)$ has to overcome the maximum value at the Primordial Nucleosynthesis to contribute with some relativistic degrees of freedom and enhance the predicted primordial abundances.\\

Resolving \eqref{chisquare} simultaneously as \eqref{omega}, it is found the following values for the free parameters:

\begin{center}\label{hohoho}
\begin{tabular}{|c|c|}\hline
 & K-essence \eqref{fun}\\ \hline
$\Omega_{\phi_0}$& 0.69 \\ \hline
$m$ & 1.0   \\ \hline
$z_*$ &  1.48 \\ \hline 
$\omega_{0}$ & - 0.99  \\ \hline 
\end{tabular}\\
\end{center}

\noindent In the figure 1 is plotted the evolution of the state equation $\omega$ as function of $a$. The field emulates radiation du\-ring this epoch and then it evolves to the next attractor: De-Sitter. Here is clear the tracker behaviour imposed in the dynamical system. In addition, it is shown that the field is relaxing to the asymptotic $\Lambda$CDM model for late times (during De-Sitter attractor.)
%

\begin{figure}[tbp]
\centering
\includegraphics[width=0.5\textwidth]{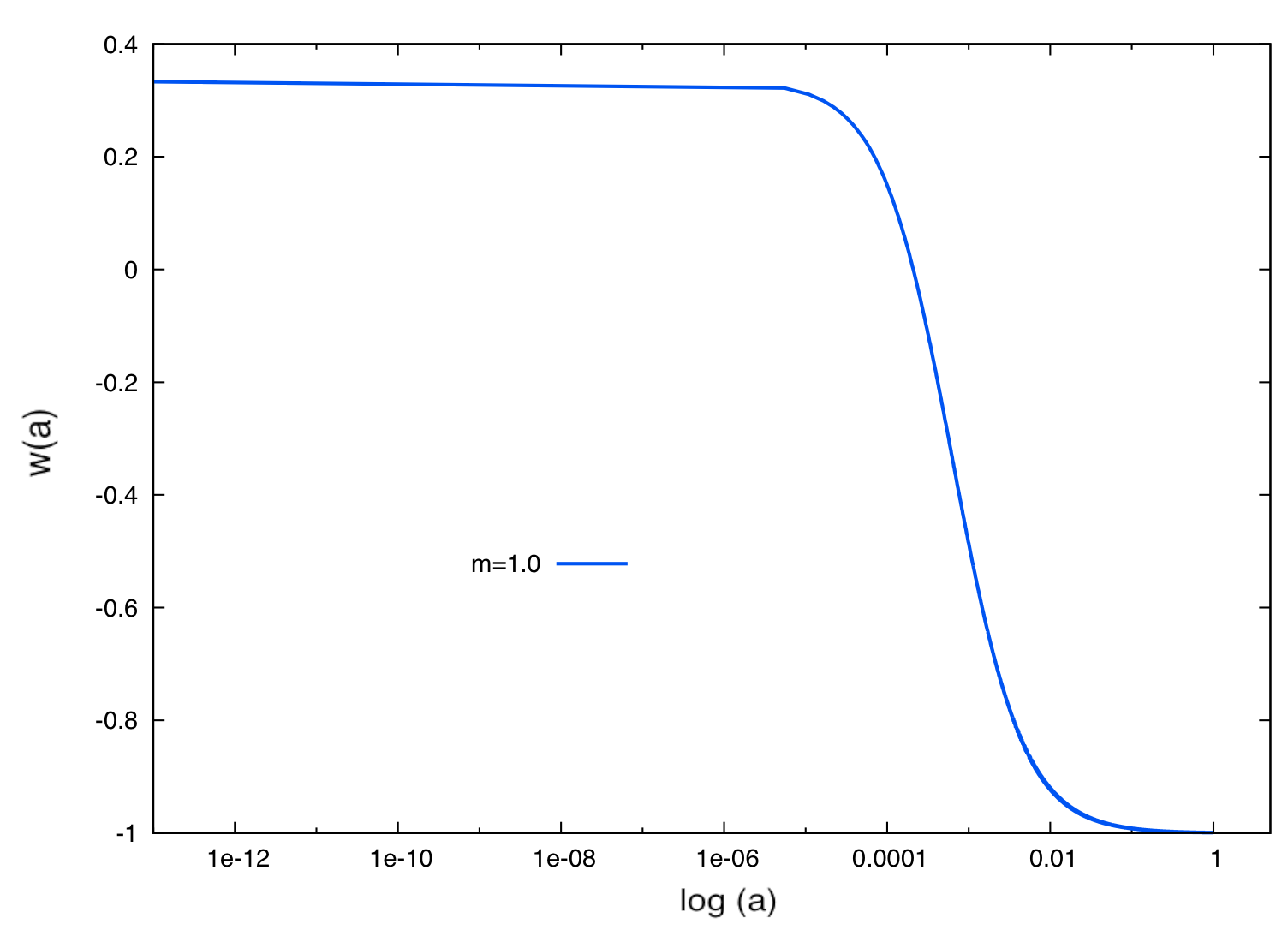}\label{Figure1}
\centering
\caption{\small{State equation of K-essence field as a function of $a$.}}
\end{figure}

Figure 2 shows the behaviour of the the function $f(z)$ that characterizes the dark energy density evolution in the model. During radiation domination epoch, the field scales as radiation $\rho \propto (1+z)^{4}$ until $z_{eq}$. After that, $\rho$ has a complex behaviour which guarantees the second attractor will be reached. At this point, the K-essence scalar field evolves in the De-Sitter attractor and its state equation goes asymptotically to $-1$ (as cosmological constant).\\

\begin{figure}[tbp]
\begin{center}
\includegraphics[width=0.5\textwidth]{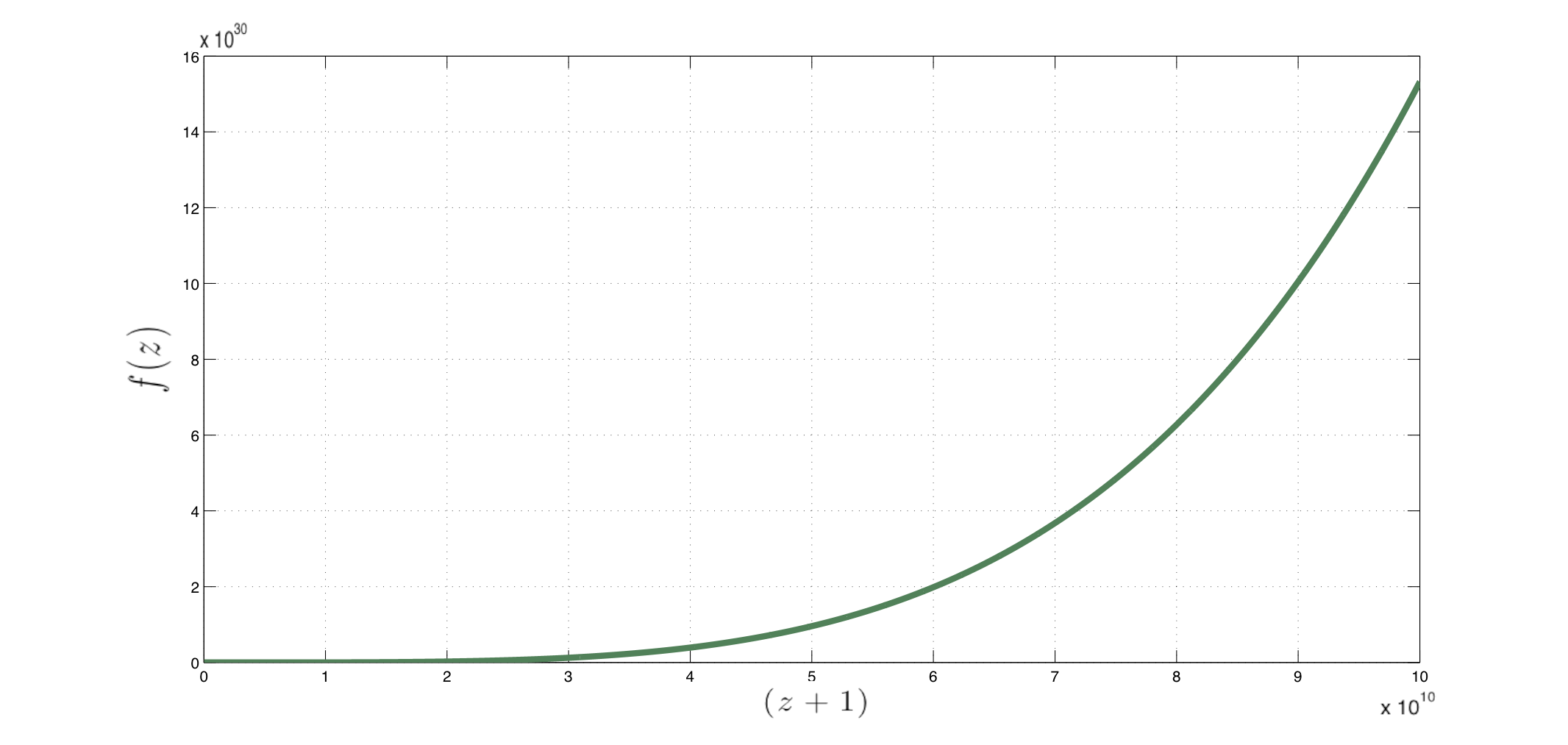} \label{Figure2}\\
\caption{\small{Function $f(z)$: dark energy density evolution.}}
\end{center}
\end{figure}

The figure 3 displays the luminosity distance for the model compared with the predicted by $\Lambda$CDM.  The shift between the curves shows that the luminosity distance is upper than the associated with the $\Lambda$CDM model, because the matter density today predicted by our model  is higher than the second one (compared with WMAP-7 $\{\Omega_{\phi 0}, \Omega_{m 0}\} = \{0.734 \pm 0.029, 0.266 \pm 0.029 \}$ \footnote{lambda.gsfc.nasa.gov/product/map/dr4/parameters.cfm}.\\

However, for low redshifts the luminosity distance grows linearly independent on the model at this regime.\\

\begin{figure}[tbp]
\begin{center}
\includegraphics[width=0.5\textwidth]{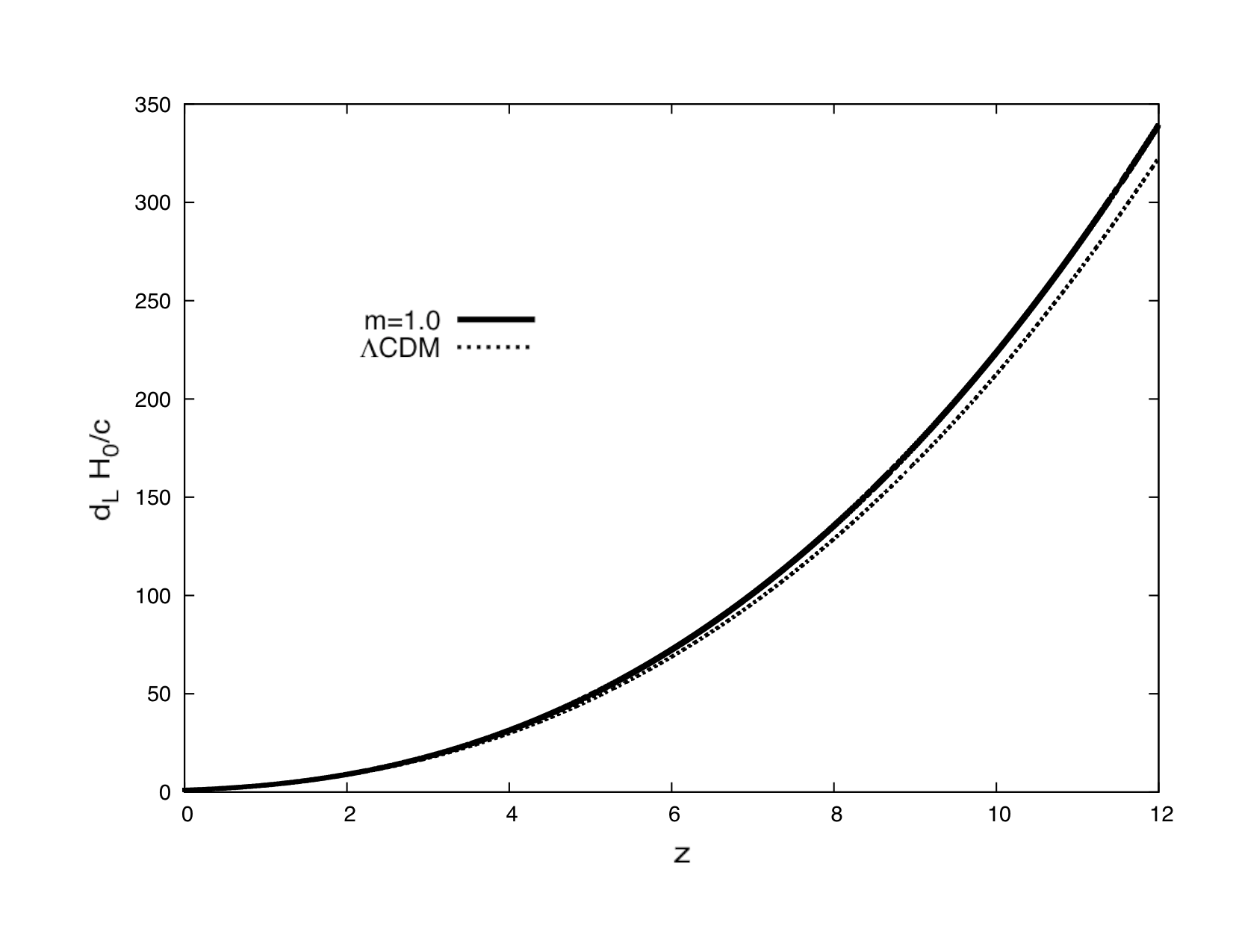}\label{Figure3}
\caption{\small{Luminosity distance as a function of $z$.}}
\end{center}
\end{figure}

On the other hand, figure 4 displays the evolution of the dark energy density fraction for the model with the estimations of the free parameters \eqref{hohoho}:

\begin{figure}[tbp]
\begin{center}
\includegraphics[width=0.5\textwidth]{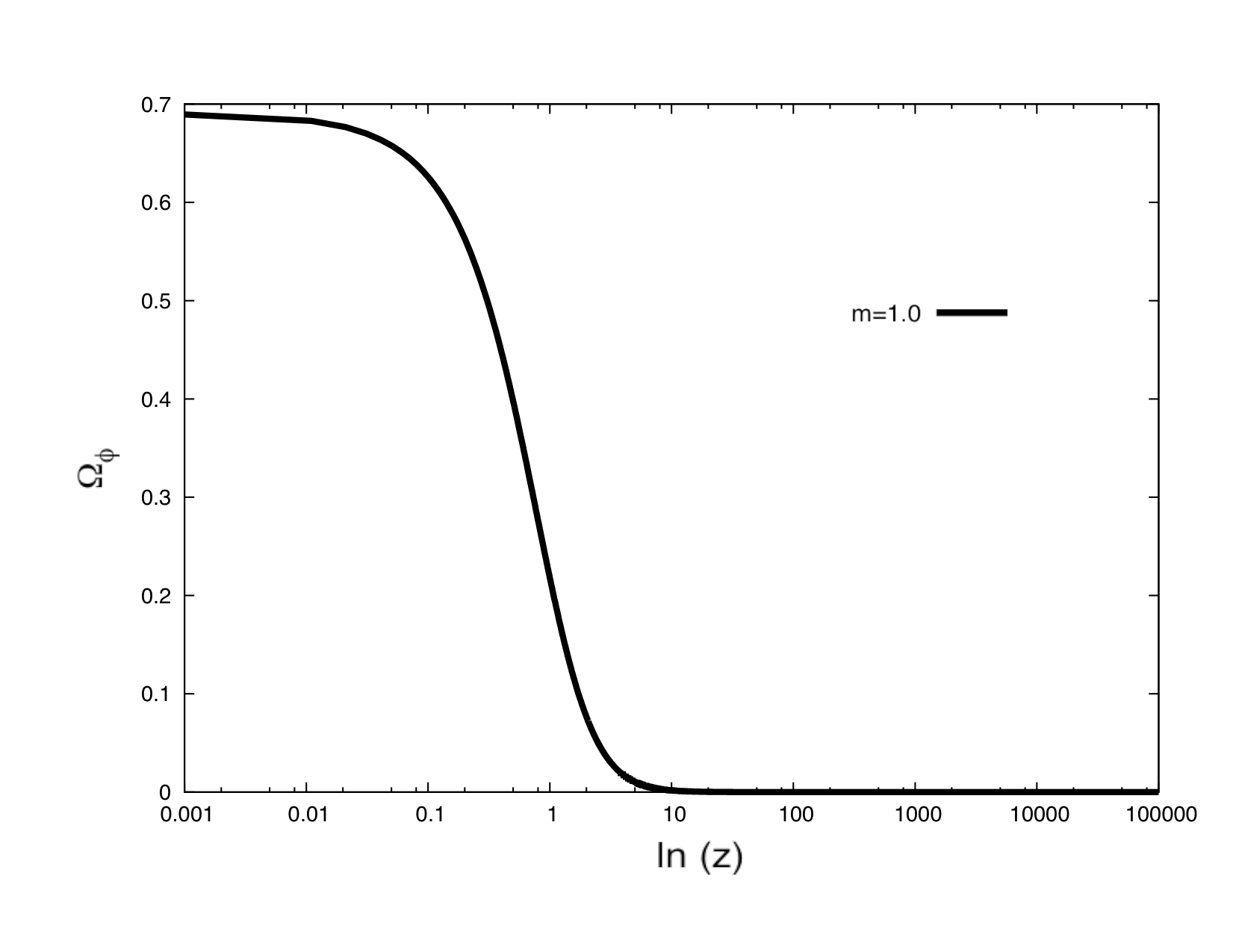}\label{Figure4} \\
\caption{\small{Evolution of the dark energy density fraction as a function of $z$.}}
\end{center}
\end{figure}

Up to now, we are interested to resolve the dynamical variables of the K-essence according to the effective parametrization we have proposed. The first one of these quantities is the adiabatic velocity of the sound $C_s^2$, which has a behaviour defined by equation \eqref{ii}.

\begin{equation}\label{ii}
C_s^2 =\omega_{\phi}+\frac{m}{4}\left(\frac{1+z_d}{1+z}\right)^m(\omega_{\phi}+1).
\end{equation}

\noindent Using \eqref{fun}, we complete the evolution of $C_s^2$. The behaviour of the adiabatic velocity is plotted in the figure 5.\\

\begin{figure}[tbp]
\begin{center}
\includegraphics[width=0.5\textwidth]{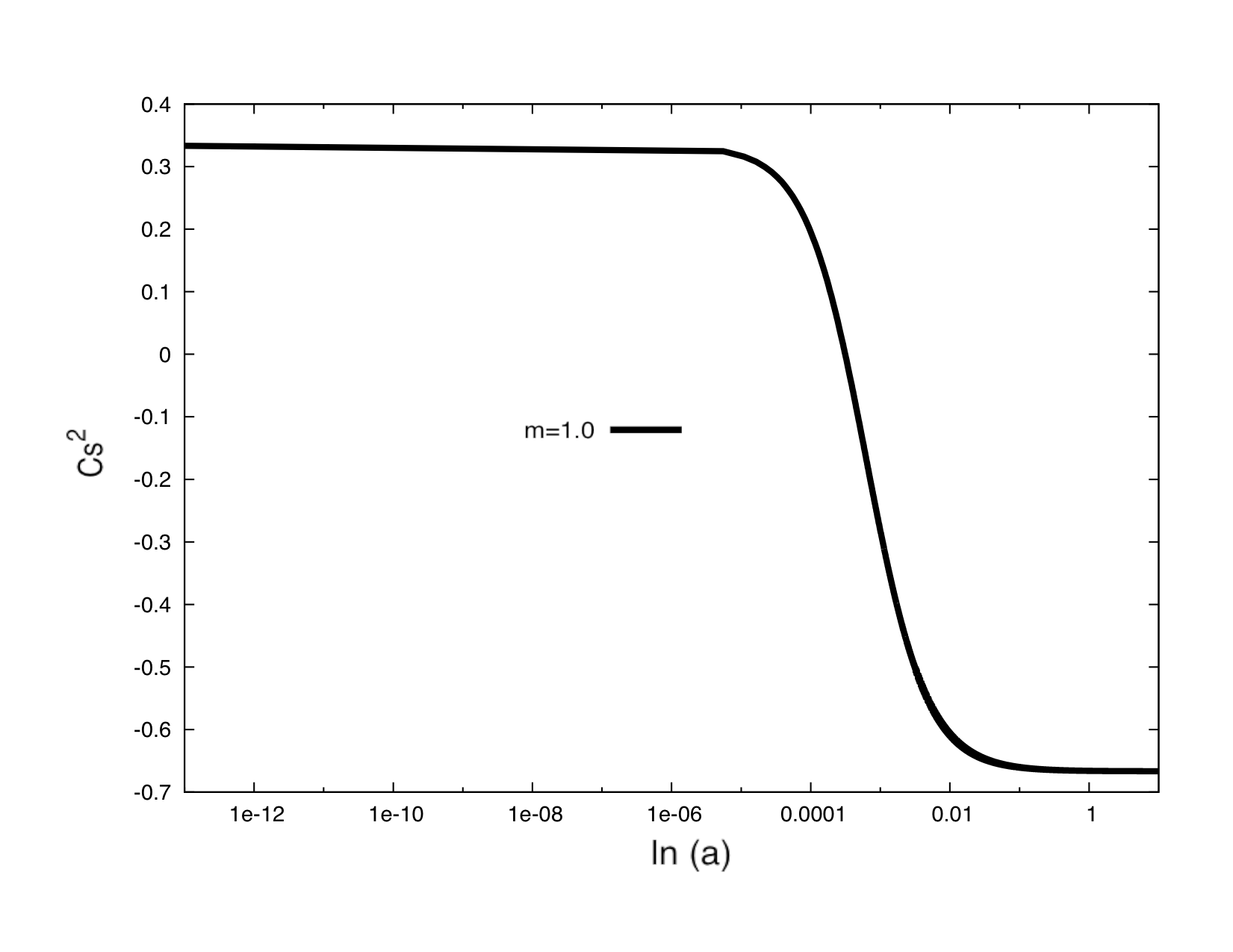}\label{Figure5} \\
\caption{\small{Adiabatic sound velocity evolution within the model.}}
\end{center}
\end{figure}

$C_s^2$ fulfills the condition \eqref{at1} in the radiation domination epoch attractor. Whatsoever, there is not a straightforward physical interpretation of the adiabatic velocity during De-Sitter attractor, because this value implies a complex Young module of the plasma perturbations.\\
It might be that the field is unstable when $C_s^2<0$, however, it is necessary to compute the evolution of the field perturbations and the time of the stability condition to conclude something in this respect.

On the other hand, the velocity of the field (the time evolution of the field) has a solution in terms on the scale factor given by:

\begin{equation}\label{vend}
v=v_0 \cdot (1+z) \left( \frac{\left(\frac{1+z_d}{1+z}\right)^m+1}{\left(1+z_d\right)^m+1}\right)^{4/m-1}.
\end{equation}

\noindent where $v_0$ is the field velocity today.

Finally, the non-canonical term of the action results:

\begin{equation}\label{qend}
Q=Q_0 \cdot (1+z)^{4} \left(\frac{\left(\frac{1+z_d}{1+z}\right)^m+1}{\left(1+z_d\right)^m+1} \right) ^{4/m+1} \left(\frac{\left(1+z_d\right)^m-1/3}{\left(\frac{1+z_d}{1+z}\right)^m-1/3} \right).
\end{equation}

\noindent That is the predicted behaviour for the non-canonical term of the lagrangian in terms on $z$. More interesting is that $Q(v)$ evolves as radiation in the first attractor and then acquires a more complicated dependence during the transition between the attractors, ensuring the continuity of the function.

\section{Standard cosmological proves}

\subsection*{Age of the universe with this model}

The age of the universe according to our model is given by the expression:

\begin{equation*}\label{age}
t_0 =\frac{1}{H_0}\int_0^1 \frac{a^{-1} da}{(0.31 a^{-3}+0.69 f(a))^{1/2}},\\
t_0 =1.2987\times 10^{10} \text{years}.
\end{equation*}

\noindent This is just an approximation of the age of the universe; however, this is an excellent result taking into account that the parametrization was figured out for $z<10^{15}$. \\\\

On the other hand, the existence of this scalar field implies that the universe evolves faster than in the standard model, because for a more negative $\omega$, more accelerated is the expansion and older is the universe for a given $H_0$. 

\subsection*{Matter inhomogeneities evolution}

When the Meszaros equation is resolved within the model with the K-essence scalar field, one solution for the inhomogeneous modes grow with the Hubble factor de Hubble. Those are the modes which maintain their amplitude after cross the horizon.\\

There are many effects on the density of perturbations: it is suppressed the linear growth $\omega(z) \leq -1/2$, with respect $\Lambda$CDM model, where the growth is proportional to the scale factor $a$; this suppression raises with higher $\omega$ at matter domination epoch and entails an earlier beginning to the dark energy domination epoch (accelerated expansion).

Otherwise, considering the CMB anisotropies spectrum, it is possible to verify that for values of $m>1.0$, there is formation of the first two acoustics peaks, then the model has a strong influence during radiation and matter domination epochs, which not correspond with the observations. On the other hand, for $m$ values lower than $0.5$, the spectrum goes faster to $\Lambda$CMB, and they are rejected because do not scale as radiation in that domination but tend $\omega_{\phi}\sim -1$ for $a<<a_{*}$. \\\\

The acoustic peaks correspond to the modes which at the CMB decoupling epoch were in the maximum compression (odds peaks) or rarefaction (even) and their position is really sensitive with the state equation of the dark energy. Actually, the first peak depends on $\Omega_{de}$ monotonically and taking into account that our model predicts a lower value for this parameter compared with the $\Lambda$CDM, hence the first peak undergoes a shifting to lower multipole moments.
However, there is a degeneracy that must be broken down in this phenomenon: the existence of this kind of model implies a rise in the number of baryons $\Omega_B$, that also entails a shifting to lower multipole moments of the peaks. Furthermore, the larger $\Omega_B$ is, the higher would be the first peak.\\ 

Therefore, if $\Omega_m$ increases its value within the model, the relative distance between the peaks decreases, because the mass associated with the baryons makes that the oscillations occur faster.

\subsection*{\textit{Statefinder} parameters}

In order to distinguish between our model and $\Lambda$CDM, \cite{rs} has proposed a test using \textit{Statefinder} parameters, defined by:

\begin{align}\label{rsprime}
r&=1+\frac{9}{2}\Omega_{\phi}\omega_{\phi}(1+\omega_{\phi})-\frac{3}{2}\Omega_{\phi}\frac{\dot{\omega_{\phi}}}{H}, \\
s&=1+\omega_{\phi}-\frac{1}{3}\frac{\dot{\omega_{\phi}}}{H\omega_{\phi}},
\end{align}

When the parameters are computed, we concluded that the K-essence scalar field model is slightly off $\Lambda$CDM, because the prediction of the values $\Omega_{m 0}$, $\Omega_{\phi 0}$ and $z_*$ are upper than the obtained with cosmological constant. However, as we have argued in this section, the model is relaxed and tends to $\Lambda$CDM for redshifts during De-Sitter domination epoch.

\subsection*{CMB shift parameter $R$}

This parameter measures the shifting of the acoustic peaks from from \textit{BAO} and it is defined as the comoving distance between the last scattering surface and today:

\begin{equation}
R=(\Omega_m H_0^2)^{1/2}\int_0^{1089}\frac{dz}{H(z)},
\end{equation}

\noindent with $H(z)$ depends strongly on the model.\\
The measured value for this parameter is $R=1.719 \pm 0.019$ \cite{panotopoulos}, meanwhile the numerical calculations made with our model give $R_{cal}=1.75$; the convergence to this value is insured for $z<200$.

\section{Constrains with BBN abundances}

As it has been proposed in the introduction of this paper, the main goal with the effective parametrization of the state equation for the K-essence scalar field is to avoid the need to fine tuning the initial conditions, but also reproduce the BBN assess enhancing the predicted abundances. The devise is really simple, we include the field energy density as relativistic degrees of freedom during radiation epoch, then compute the Hubble factor and consequently the Boltzmann equations for the light nuclei.\\

Firstly, we set out a fiducial baryon to photon ratio $\eta=6.2\times 10^{-9}$ \footnote{lambda.gsfc.nasa.gov/product/map/dr4/parameters.cfm} and calculate the capture temperature that deuterium began to be formed via nuclear reactions given by:

\begin{equation}\label{Tcapture}
T_{c,\gamma}=\epsilon_D/26=0.088 MeV.
\end{equation}

For this temperature the deuterium is not longer photodissociate \cite{bernstein2}, where $\epsilon_D$ is the deuterium binding energy. Temperatures lower than $T_{c;\gamma}$ the whole chains related with primordial nucleosynthesis could be run.\\

Following the discussion of Bernstein , we assess analytically the value of the mass fraction $Y_{^4He}$ and then, using \textit{FastBBN} \cite{kawano, kellogg}, \textit{Public BBN} and \textit{BBNreactions} \cite{bbnreactions}, we compute numerically the light abundances.\\

In order to introduce the effective degrees of freedom of the K-essence scalar field, we impose the condition mentioned in section III to restrict the value of the m-parameter: we want to achieve the largest contribution of the energy density, been subdominant with respect to the radiation energy density. The condition can be quantified as:

\begin{equation}\label{hungry}
\rho_{\phi}\vert_{rad}=b \cdot \rho_{rad} \:\:\:\:\:\:\:\:\:\:\:\:\:\:\: 0\leq b < 1.
\end{equation}

We introduce this energy density in the Hubble parameter during radiation era and executed the time capture, the neutron's fraction at this time and finally, the mass fraction, obtaining the following results:

\begin{center}\label{Table a}
\begin{tabular}{|c|c|c|}\hline 
& $b=0$ & \tiny{Model with a contribution of $b=0.2$} \\ \hline
\tiny{$H(b)$ ($s^{-1}$)} & 1.13  & 1.2379 \\ \hline
\tiny{$t_c$ ($s$)} & 182 &  169.8  \\ \hline 
\tiny{ $X_n(t_c)$} &0.123 & 0.125   \\ \hline
\tiny{$Y_4$}& 0.247 & 0.249 \\ \hline
\end{tabular}\\
\end{center}

The values in the \ref{Table a} are compared with the calculated values within the standard model (where $b=0$, \textbf{i.e.} null contribution of the K-essence scalar field).\\

The abundances for the other nuclei are not precisely computed to be reported using this analytical method, because they depend on the coupled Boltzmann chains.\\

\noindent However, it is notable that the predicted values for the $^4He$ is in agreed with the observational boundaries, therefore our model is an excellent candidate for dy\-na\-mi\-cal dark energy model in FRWL.\\

Using \textit{FastBBN} \cite{kawano, kellogg}, \textit{Public BBN} and \textit{BBNreactions} \cite{bbnreactions}, we compute numerically the light abundances including the relativistic degrees of freedom of the field as $b$:

\begin{widetext}
\begin{center}\label{Table b}
\begin{tabular}{|c|c|c|c|c|c|}\hline
C\'ode & $b$ & $D/H \times 10^{-5}$ &  $^3He/H \times 10^{-5}$ &$Y_p$ &  $^7Li/H \times 10^{-10}$ \\  \hline
\tiny{Fast BBN} & $0.0$ & $2.335$ &$1.546$ & $0.241$ & $1.268$ \\ 
\tiny{Fast BBN} &$5.0\times 10^{-3}$ & $2.345$&$1.548$&$0.241$& $1.261$\\   
\tiny{Fast BBN} &$5.0\times 10^{-2}$ & $2.436$&$1.570$ & $0.245$ & $1.263$ \\  
\tiny{Fast BBN} &$0.1$ & $2.537$ &$1.594$  & $0.249$ & $1.261$ \\  
\tiny{Fast BBN} &$0.2$ & $2.741$&$1.639$ & $0.256$ & $1.269$\\ \hline    
\tiny{BBN reactions}& $0.0$ & $4.043$ & $2.363$ & $0.243$ & $1.543$\\ \hline
\tiny{Public Big bang} & $0.0$& $1.542$ &$3.000$ & $0.242$ & $4.884$ \\ 
\tiny{Public Big bang} &$5.0\times 10^{-3}$ & $1.547$ &$3.002$  & $0.242$ & $4.907$ \\ 
\tiny{Public Big bang} &$5.0\times 10^{-2}$ & $1.5845$ &$3.015$ & $0.246$ & $5.111$\\ 
\tiny{Public Big bang} &$0.1$ & $1.629$ &$3.030$  & $0.250$ &$5.345$ \\ 
\tiny{Public Big bang} &$0.2$ &$1.717$ &$3.059$ & $0.257$ & $5.825$\\ \hline  
\tiny{our EDE model}& $0.2$& -- & -- & $0.249$& -- \\ \hline
\cite{wmap1}&$0.0$& $2.75\pm 0.24$ &$0.93 \pm 0.055$ & $0.2484\pm 0.0004$ & $3.82 \pm 0.66$ \\ \hline
\cite{wmap2}&$0.0$&$2.60\pm 0.18$ &$1.04 \pm 0.04$ & $0.2479\pm 0.0004$ & $4.15 \pm 0.47$ \\ \hline
\end{tabular}\\
\end{center}
\end{widetext}

Moreover, it is possible to assess the baryon to photon ratio from the CMB temperature anisotropies spectrum, taking into account the relation:

\begin{equation}
\eta_B=\frac{n_B}{n_{\gamma}}=5.5\times 10^{-10}\left(\frac{\Omega_B h^2}{0.022}\right)
\end{equation}

\noindent In the table 4 are shown the values of the parameters related with BBN ($\Omega_B $ and $\eta_B$)  in terms of $b$, where $b=0.0$ is the associated value to $\Lambda$CDM and $b=0.2$, the maximum contribution of the field during radiation domination epoch:

\begin{center} \label{Table c}
\begin{tabular}{|c|c|c|c|}\hline
 $\rho_{\phi}$ &  $\Omega_B $ &$\Omega_B h^2$ & $\eta_B \times 10^{-10}$  \\ \hline
$b=0.0$ &  0.044 & 0.02218 & 6.20 \\ \hline
$b=0.2$ &  0.053 &0.02692 & 6.73 \\ \hline
\cite{wmap1} &  \tiny{$0.0449 \pm 0.0028$}& \tiny{$0.02258 \pm 0.00057$} & \tiny{$6.190 \pm 0.145$}   \\ \hline
\end{tabular}\\
\end{center}

According to the results shown in the table \ref{Table c}, we plot the abundances for the light nuclei including $^4He$, for different contribution of the field, implemented as effective relativistic degrees of freedom in the code \textit{Public BBN}.

\begin{figure}[tbp]
\begin{center}
\includegraphics[width=0.5\textwidth]{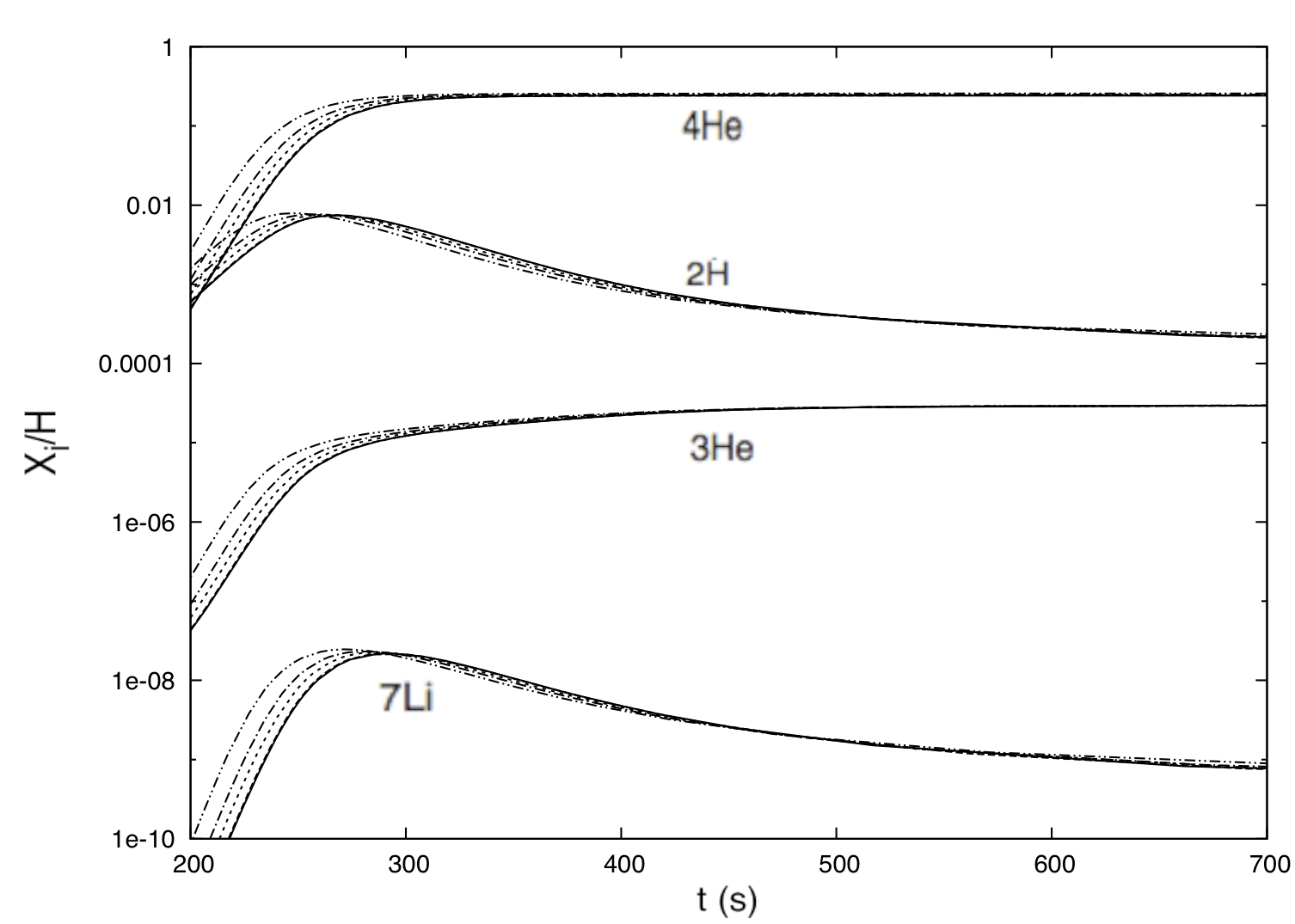}\label{Figure6}
\caption{\small{$b=0.0$ (solid line), $b=5.0\times 10^{-3}$ (dashed line), $b=5.0\times 10^{-2}$ (dotted line), $b=0.1$ (width line) and $b=0.2$ (dashed dotted line).}}
\end{center}
\end{figure}

\section{Conclusions}

In this work, we have made a general description of the dark energy component as a K-essence scalar field that evolves during radiation as a tracker ($\omega_{\phi}\vert_{rad}=\frac{1}{3}$) and at late times achieves the De-Sitter attractor $\omega_{\phi} \rightarrow -1$, emulating cosmological constant $\Lambda$. All the description was made in the hot cosmological plasm including matter-radiation components and assuming a spacially flat universe. With these assumptions for the field, we have proposed an effective parametrization for the state equation $\omega_{\phi}$ \eqref{fun}, that its free parameters estimations were obtained by minimizing the function $\chi^2$ with the distances modulus of the Type Ia Supernovae.\\

We have rewritten the velocity of the field and the non-canonical kinetic term of the K-essence lagrangian in terms of $z$, obtaining the completed behaviour of the field during the thermal history and this let us say that K-essence system includes different classes of Quintessence, when it is considered specific cases of the non-canonical term. Another advantage with this formulation is that avoids the \textit{Fine-tuning} of the initials conditions of the field and its velocity.\\
As it was expected, during radiation and matter domination epochs the field has upper predictions for some observables $\Omega_{m 0}$, $\Omega_{\phi 0}$ and $z_*$ (but respecting the observational bounds for them) comparing with $\Lambda$CDM model. However, at late times $z\sim 0$ tends asymptotically the standard model, after it has evolved from the radiation  to the second attractor. All these results are in agreement with the conditions that have been imposed to resolve the dynamical system \cite{kang}.\\
On the other hand, when it is included the non-null contribution of the field during radiation domination era, the Hubble factor is affected, but also the time capture and therefore, the primordial light nuclei abundances, because there were more neutrons out of the equilibrium to form $^4$He by two body-reactions. Actually, the whole reactions occurred faster, such that the production of the nuclei are more effective and drives in an upper mass fraction.\\

The predicted value for the $^4$He abundance prediction according to our model is inside the observational bounds, then, our model is an excellent candidate to a dynamic dark energy with a subdominant contribution during radiation and matter epoch.\\
Finally, it is remarkable that the results of this paper can be compared with other kind of models, because the field degrees of freedom can be treated as effective degrees of some other component (for instance, a Quintessence scalar field or even, sterile neutrinos \cite{lua}). In fact, we have include the field contribution in the numerical codes in these way, and it let us to compare degenerations of different kind of models.\\

\begin{acknowledgements}
This work is supported and developed by the Observatorio Astron\'omico Nacional under the auspices of Universidad Nacional de Colombia. \\ 
We also want to thank Daniel Molano, Carlos Cede\~no and all the professors who give us an advise to enhance this paper in the international events that this work was presented.
\end{acknowledgements}

\end{document}